\documentstyle[12pt]{article}

\advance \topmargin by -\headheight
\advance \topmargin by -\headsep

\evensidemargin \oddsidemargin
\def\ie{{\em i.e.}}

\def\ie{\hbox{\it i.e.}}

\def\CC{{\mathchoice
{\rm C\mkern-8mu\vrule height1.45ex depth-.05ex 
width.05em\mkern9mu\kern-.05em}
{\rm C\mkern-8mu\vrule height1.45ex depth-.05ex 
width.05em\mkern9mu\kern-.05em}
{\rm C\mkern-8mu\vrule height1ex depth-.07ex 
width.035em\mkern9mu\kern-.035em}
{\rm C\mkern-8mu\vrule height.65ex depth-.1ex 
width.025em\mkern8mu\kern-.025em}}}

\def\RR{{\rm I\kern-1.6pt {\rm R}}}

\def\ZZ{{\rm Z}\kern-3.8pt {\rm Z} \kern2pt}

\def\np{Nucl. Phys.}
\def\pl{Phys. Lett.}

\def\pr{Phys. Rev.}

\def\jhep{J. High Energy Phys.}

\newcommand{\beq}{\begin{equation}}
\newcommand{\eeq}{\end{equation}}
\newcommand{\rc}{\nonumber\\}
\newcommand{\bear}{\begin{eqnarray}}
\newcommand{\eear}{\end{eqnarray}}

\newfont{\namefont}{cmr10}
\newfont{\addfont}{cmti7 scaled 1440}
\newfont{\boldmathfont}{cmbx10}
\newfont{\headfontb}{cmbx10 scaled 1728}
\renewcommand{\theequation}{{\rm\thesection.\arabic{equation}}}
\begin{document}
\begin{titlepage}

\begin{center} \Large \bf Giant Gravitons with NSNS B field

\end{center}

\vskip 0.3truein
\begin{center} 
J. M. Camino
\footnote{e-mail:camino@fpaxp1.usc.es}
and 
A.V. Ramallo
\footnote{e-mail:alfonso@fpaxp1.usc.es}

\vspace{0.3in}

Departamento de F\'\i sica de
Part\'\i culas, \\ Universidad de Santiago\\
E-15706 Santiago de Compostela, Spain. 
\vspace{0.3in}

\end{center}
\vskip 1truein

\begin{center}
\bf ABSTRACT
\end{center} 

We study the motion of a D(8-p)-brane probe in the background created by a
stack of non-threshold (D(p-2), Dp) bound states for $2\le p\le 6$. The brane
probe and the branes of the background have two common directions. We show that
for a particular value of the worldvolume gauge field there exist configurations
of the probe brane which behave as  massless particles and can be interpreted as
gravitons blown up into a fuzzy sphere and a noncommutative plane. We check this
behaviour by studying the motion and energy of the brane and by determining how
supersymmetry is broken by the probe as it moves under the action of the
background.

\vskip4.5truecm
\leftline{US-FT-11/01\hfill July 2001}
\leftline{hep-th/0107142}
\smallskip
\end{titlepage}
\setcounter{footnote}{0}

%%%%%%%%%%%%%%%%%%%%%%%%%%%%%%%%%%%%%%%%%%%%%%%%%%%%%%%%%%%%%%%%%%%%%%%%%%%%%%
%%%%%          		    M A I N   T E X T 
%%%%%%%%%%%%%%%%%%%%%%%%%%%%%%%%%%%%%%%%%%%%%%%%%%%%%%%%%%%%%%%%%%%%%%%%%%%%%%
\setcounter{equation}{0}
\section{Introduction}
\medskip

One of the most interesting things we have recently learnt from string theory is
the fact that a system can increase its size with increasing momentum. There are
several manifestations of this phenomenon, which is opposite to the standard
field theory intuition, in several contexts related to string theory such as  the
infrared ultraviolet connection \cite{UVIR} or the noncommutative geometry
\cite{NCG}.

In ref. \cite{GST} McGreevy, Susskind and Toumbas found another example of the
growth in size with energy. These authors considered a massless particle moving
in a spacetime of the form $AdS_m\times S^{p+2}$ and discovered that there
exists a configuration in which an expanded brane (the giant graviton) has
exactly the same quantum numbers as the point-like particle. This expanded brane
wraps the spherical part of the spacetime and is stabilized against shrinking
by the flux of the Ramond-Ramond (RR) gauge field. The size of the giant graviton
increases with its angular momentum and, as the radius of the brane cannot be
greater than the radius of the spacetime, one gets that there exists an upper
bound for the momentum of the brane. This fact is a realization of the
so-called stringy exclusion principle. Moreover, in refs. \cite{GMT, HHI}  was
proved that the giant gravitons of ref. \cite{GST} are BPS configurations which
preserve the same supersymmetry as the point-like graviton. It was also shown in
\cite{GMT, HHI} that there also exist gravitons expanded into the $AdS$ part of
the spacetime which, however, do not have an upper bound on their angular
momentum due to the non-compact nature of the $AdS$ spacetime. 

The overall physical picture one gets from these results is that for high
momenta the linearized approximation to supergravity breaks down and one is
forced to introduce interactions in order to describe the dynamics of the 
massless modes  of the theory. An effective procedure to represent these
interactions is to assume that the massless particles polarize and become a
brane. The precise mechanism of this polarization is the so-called Myers
dielectric brane effect \cite{Myers}. 

The blowing up of gravitons into branes can take place in backgrounds different
from $AdS_m\times S^{p+2}$. Indeed, in ref. \cite{DTV} it was found that there
are giant graviton configurations of D(6-p)-branes moving in the near-horizon
geometry of a dilatonic background created by a stack of Dp-branes. For other
aspects of the expanded graviton solutions see refs. \cite{DJM}-\cite{KM}.

In this paper we will find giant graviton solutions for probes which move in
the geometry created by a stack of non-threshold bound states of the type 
(D(p-2), Dp) for $2\le p \le 6$  \cite{NCback}. These backgrounds 
are $1/2$ supersymmetric and have been
considered recently \cite{MR} as supergravity duals of noncommutative gauge
theories. They are characterized by the presence of a non-zero Kalb-Ramond field
$B$,  directed along the Dp-brane,  from the Neveu-Schwarz sector of the
superstring, together with the corresponding RR fields. We will place in this
background a brane probe in such a way that it could capture both the RR flux
(as in the Dp backgrounds) and the flux of the $B$ field. This last requirement
implies that we must extend our probe along two directions parallel to the
background and, actually, our probe will be a D(8-p)-brane wrapped on an
$S^{6-p}$ sphere transverse to the background and extended along a
(noncommutative) plane parallel to it. We will verify that, by switching on a
particular worldvolume gauge field, one can find configurations of the
D(8-p)-brane which behave as a massless particle.  We will check that the
energy  of these giant graviton configurations is exactly the same as that of a
massless particle moving in the metric of the (D(p-2), Dp) background.
Generically, the brane falls into the center of the gravitational potential
along a trajectory which will be determined. We will also study the supersymmetry
projection introduced by the brane and we will show that it breaks completely
the supersymmetry of  the background, exactly in the same way 
as  a wave which propagates with the velocity of the
center of mass of the brane. Thus, also from the point of view of supersymmetry,
the expanded brane mimics the massless particle. 

The plan of this paper is the following. In section 2 we will
start  our analysis by  describing the (D(p-2),Dp) background. Then, we will
consider a brane probe and  its action and energy on the background will be
studied. At the end of this section the giant graviton configurations will be
characterized and the corresponding equations of motion will be integrated.
Section 3 is devoted to the analysis of the supersymmetry of the problem for the
particular case of the (D1,~D3) background. First of all, we shall obtain the
form of the Killing spinors. Then, the $\kappa$-symmetry of the probe is studied
and its associated  supersymmetry projection is obtained. Details of the
determination of the Killing spinors for the (D1,~D3) background are given in
appendix A. Finally, in section 4 we will summarize  our results and discuss
some possible extensions of our work.

\medskip
\setcounter{equation}{0}
\section{Giant gravitons in (D(p-2),Dp)  backgrounds}
\medskip

The supergravity background we will consider is the one generated  by a stack of
$N$  non-threshold bound states of Dp and D(p-2) branes for 
$2\le p \le 6$. The metric and dilaton (in the string frame) for such a
background are \cite{NCback}:
\bear
ds^2&=&f_p^{-1/2}\,\Big[\,-(\,dx^0\,)^2\,+\,\cdots\,+\,(\,dx^{p-2}\,)^2\,+\,
\,h_p\,\Big((\,dx^{p-1}\,)^2\,+\,(\,dx^{p}\,)^2\Big)\,\Big]\,+\rc\rc
&&+\,f_p^{1/2}\, \Big[\,dr^2\,+\,r^2\,d\Omega_{8-p}^2\,\Big]\,\,,\rc\rc
e^{\tilde\phi_D}&=&f_p^{{3-p\over 4}}\,\,h_p^{1/2}\,\,,
\label{uno}
\eear
where $d\Omega_{8-p}^2$ is the line element of a unit $(8-p)$ sphere, 
$r$ is a radial coordinate parametrizing the distance to the brane bound state
and $\tilde\phi_D=\phi_D\,-\,\phi_D(r\rightarrow\infty)$. The Dp-brane of the
background extends along the directions $x^0\cdots x^p$, whereas the 
D(p-2)-brane component lies along $x^0\cdots x^{p-2}$. The functions $f_p$ and
$h_p$ appearing in (\ref{uno}) are:
\bear
f_p&=&1\,+\,{R^{7-p}\over r^{7-p}}\,\,,\rc\rc
h_p^{-1}&=&\sin^2\varphi\,f_p^{-1}\,+\,\cos^2\varphi\,\,,
\label{dos}
\eear
with $\varphi$ being an angle which characterizes the degree of mixing of the 
Dp and D(p-2) branes in the bound state. The parameter $R$, which we will refer
to as  the radius,   is given by:
\beq
R^{7-p}\,\cos\varphi\,=\,N\,g_s\,2^{5-p}\,\pi^{{5-p\over 2}}\,
(\,\alpha\,'\,)^{{7-p\over 2}}\,\,
\Gamma\Bigl(\,{7-p\over 2}\Bigr)\,\,,
\label{tres}
\eeq
where $N$ is the number of branes of the stack, $g_s$ is the string coupling
constant ($g_s=e^{\phi_D(r\rightarrow\infty)}$) and $\alpha\,'$ is the Regge
slope.

This solution is also endowed with a rank two  NSNS field $B$  directed
along the $x^{p-1}x^{p}$ (noncommutative) plane:
\beq
 B=\tan\varphi\,f_p^{-1}\,h_p\,\,dx^{p-1}\wedge dx^{p}\,\,,
\label{cuatro}
\eeq
and is charged under RR field strengths,  $F^{(p)}$ and 
$F^{(p+2)}$. The components along the directions parallel to the bound state
are:
\bear
F^{(p)}_{x^0,x^1,\cdots, x^{p-2},r}
&=&\sin\varphi\,\partial_r\,f_p^{-1}\,\,,\rc\rc
F^{(p+2)}_{x^0,x^1,\cdots, x^{p},r}
&=&\cos\varphi\,
h_p\partial_r\,f_p^{-1}\,.
\label{cinco}
\eear
It is understood that the  $F^{(p)}$'s for $p\ge 5$ are the
Hodge duals of those with $p\le 5$, \ie\ $F^{(p)}={}^*F^{(10-p)}$ for $p\le 5$.
In particular this implies that $F^{(5)}$ is self-dual, as is well-known for
the type IIB theory. It is clear from the above equations that for $\varphi=0$
the (D(p-2),Dp) solution reduces to the Dp-brane geometry whereas for 
$\varphi=\pi/2$ it becomes a D(p-2)-brane smeared along the $x^{p-1}x^p$
directions.

It will be convenient for our purposes to use a particular coordinate system
\cite{GST} for the unit $8-p$ sphere $S^{8-p}$. In order to define these
coordinates, let us realize  $S^{8-p}$ as the surface in $\RR^{9-p}$ whose
equation is 
$(\,z^1\,)^2\,+\cdots+\,(\,z^{9-p}\,)^2\,=\,1$. We shall parametrize
$z^1$ and $z^2$ by means of the   coordinates  $\rho$ and $\phi$ as follows:
\beq
z^1\,=\,\sqrt{1\,-\,\rho^2}\,\cos\phi\,\,,
\,\,\,\,\,\,\,\,\,\,\,\,\,\,\,\,\,\,
z^2\,=\,\sqrt{1\,-\,\rho^2}\,\sin\phi\,\,,
\label{seis}
\eeq
where $0\le\rho\le 1$ and $0\le\phi\le 2\pi$. Clearly, on $S^{8-p}$, one has:
\beq
(\,z^1\,)^2\,+\,(\,z^2\,)^2\,=\,1-\rho^2\,\,,
\,\,\,\,\,\,\,\,\,\,\,\,\,\,\,\,\,\,
(\,z^3\,)^2\,+\cdots+\,(\,z^{9-p}\,)^2\,=\,\rho^2\,\,.
\label{siete}
\eeq
After a simple calculation one can demonstrate that, in terms of $(\rho,\phi)$, 
the metric of $S^{8-p}$ takes the form:
\beq
d\Omega_{8-p}^2\,=\,{1\over 1-\rho^2}\,d\rho^2\,+\,
(1\,-\,\rho^2\,)\,d\phi^2\,+\,\rho^2\,d\Omega_{6-p}^2\,\,,
\label{ocho}
\eeq
where $d\Omega_{6-p}^2$ is the metric of a unit $6-p$ sphere.

The Hodge duals of the RR field strengths can be easily computed from eqs. 
(\ref{cinco}) and (\ref{uno}). Clearly ${}^*\,F^{(p)}$ is a 
$(10-p)$-form whereas ${}^*\,F^{(p+2)}$ is a $(8-p)$-form. After a simple
calculation one can check that  ${}^*\,F^{(p)}$ and ${}^*\,F^{(p+2)}$
have  the following components:
\bear
{}^*\,F^{(p)}_{x^{p-1},x^p,\rho,\phi,\theta^1\cdots, \theta^{6-p}}
&=&(-1)^{p+1}\,(7-p)\,\sin\varphi\,R^{7-p}\,\rho^{6-p}\,
h_p\,f_p^{-1}\,\sqrt{\hat g^{(6-p)}}\,\,,\rc\rc
{}^*\,F^{(p+2)}_{\rho,\phi,\theta^1\cdots,\theta^{6-p}}
&=&(-1)^{p+1}\,(7-p)\,\cos\varphi\,R^{7-p}\,\rho^{6-p}\,
\,\sqrt{\hat g^{(6-p)}}\,\,,
\label{nueve}
\eear
where $\theta^1,\cdots, \theta^{6-p}$ are coordinates of the unit $6-p$ sphere 
and $ \hat g^{(6-p)}$ is the determinant of the $S^{6-p}$ metric. These forms 
satisfy the equations:
\beq
d{}^*\,F^{(p)}\,=\,H\,\wedge\,{}^*\,F^{(p+2)}\,\,,
\,\,\,\,\,\,\,\,\,\,\,\,\,\,\,\,\,\,
d{}^*\,F^{(p+2)}\,=\,0\,\,.
\label{diez}
\eeq
Then, one can represent ${}^*\,F^{(p)}$ and ${}^*\,F^{(p+2)}$ in terms of
three potentials as follows:
\bear
{}^*\,F^{(p)}&=&dC^{(9-p)}\,-\,H\,\wedge\,C^{(7-p)}\,\,,\rc
{}^*\,F^{(p+2)}&=&dC^{(7-p)}\,-\,H\,\wedge\,C^{(5-p)}\,\,,
\label{once}
\eear
where $H=dB$. In eq. (\ref{once}) $C^{(r)}$ is an $r$-form. Actually only for
$p=3$ the term $H\,\wedge\,C^{(5-p)}$ in the second of these equations gives a
non-vanishing contribution. By a direct calculation one can check that the other
two potentials $C^{(7-p)}$ and  $C^{(9-p)}$ have the components:
\bear
C^{(7-p)}_{\phi,\theta^1\cdots,\theta^{6-p}}
&=&(-1)^{p+1}\,\cos\varphi\,R^{7-p}\,\rho^{7-p}\,
\sqrt{\hat g^{(6-p)}}\,\,, \rc\rc
C^{(9-p)}_{x^{p-1},x^p,\phi,\theta^1\cdots, \theta^{6-p}}
&=&(-1)^{p+1}\,\sin\varphi\,R^{7-p}\,\rho^{7-p}\,
h_p\,f_p^{-1}\,\sqrt{\hat g^{(6-p)}}\,\,.
\label{doce}
\eear
Apart from the ones displayed in eqs. (\ref{cinco}) and (\ref{nueve}), 
the RR field strengths and their duals have another components. Actually, due
to the identification of $F^{(p)}$ and ${}^*F^{(10-p)}$, the forms appearing in
these equations are not all different. Using this fact it is not difficult to
obtain all the components of the  RR gauge forms. From this result one can
check that eq. (\ref{diez}) is satisfied and, thus, the representation of 
${}^*F^{(p)}$ and ${}^*F^{(p+2)}$ in terms of the different potentials appearing
on the right-hand side of eq. (\ref{once}) holds (although these potentials
have another components in addition to the ones written in eq. (\ref{doce})). 
Let us specify this for $p=3$. In this case one can easily check that $F^{(3)}$
and $F^{(5)}$ are given by:
\bear
F^{(3)}\,&=&\,\sin\varphi\,\partial_rf_3^{-1}\,
dx^0\wedge dx^1\wedge dr\,\,,\rc\rc
F^{(5)}\,&=&\,\cos\varphi\,\Big[\,h_3\partial_rf_3^{-1}
dx^0\wedge \cdots\wedge dx^{3}\wedge dr\,+\,
4R^4\rho^3d\rho\wedge d\phi\wedge\epsilon_{(3)}\,\Big]\,\,.
\label{trece}
\eear
Moreover, ${}^*F^{(3)}$ and ${}^*F^{(5)}=F^{(5)}$ can be represented in terms of
three RR potentials $C^{(6)}$, $C^{(4)}$ and $C^{(2)}$ as in eq. 
(\ref{once}) with:
\bear
C^{(6)}&=&\sin\varphi\, R^4\,\rho^4\,h_3f_3^{-1}\,
dx^2\wedge dx^2 \wedge d\phi \wedge \epsilon_{(3)}\,\,, \rc\rc 
C^{(4)}&=&\cos\varphi\,\Big[\,h_3f_3^{-1}dx^0\wedge \cdots\wedge dx^{3}\,+\,
R^4\rho^4d\phi\wedge\epsilon_{(3)}\,\Big]\,\,,\rc\rc
C^{(2)}&=&-\sin\varphi f_3^{-1} dx^0\wedge dx^1\,\,,
\label{catorce}
\eear
where $\epsilon_{(3)}$ is the volume form of the unit $S^3$ 
(with this election of $C^{(2)}$ one has $F^{(3)}\,=\,-dC^{(2)}$).  One
can treat similarly the other cases.

\medskip
\subsection{The brane probe}
\medskip

Let us now embed a D(8-p)-brane in the near-horizon region of the (D(p-2), Dp)
geometry. In this region $r$ is small and one can approximate the harmonic
function $f_p$ appearing in the supergravity solution as:
\beq
f_p\,\approx\,{R^{7-p}\over r^{7-p}}\,\,.
\label{quince}
\eeq
The D(8-p)-brane probe we will be dealing with wraps the (6-p) transverse
sphere and  extends along  the $x^{p-1}x^p$ directions. The response of the
probe to the background is determined by its action $S$, which is the sum of a
Dirac-Born-Infeld and a Wess-Zumino term \cite{swedes}:
\beq
S\,=\,S_{DBI}\,+\,S_{WZ}\,\,.
\label{dseis}
\eeq
The Dirac-Born-Infeld  term $S_{DBI}$ is:
\beq
S_{DBI}\,=\,
-T_{8-p}\,\int d^{\,9-p}\xi\,e^{-\tilde\phi_D}\,
\sqrt{-{\rm det}\,(\,g\,+\,{\cal F}\,)}\,\,,
\label{dsiete}
\eeq
where $g$ is the induced worldvolume metric,  $T_{8-p}$ is the tension of the
D(8-p)-brane:
\beq
T_{8-p}\,=\,(2\pi)^{p-8}\,(\,\alpha\,'\,)^{{p-9\over 2}}\,
(\,g_s\,)^{-1}\,\,,
\label{docho}
\eeq
and, if $P[\cdots]$ denotes the pullback to the worldvolume of a bulk field, 
${\cal F}$ is given by:
\beq
{\cal F}\,=\,F\,-\,P[B]\,=\,dA\,-\,P[B]\,\,,
\label{dnueve}
\eeq
with $F$ being a $U(1)$ worldvolume gauge field strength and $A$ its
potential. The Wess-Zumino term of the action $S_{WZ}$ couples the probe to the
RR potentials of the background. For the brane probe configuration we are
considering and the (D(p-2), Dp) background, $S_{WZ}$ is given by:
\beq
S_{WZ}\,=\,T_{8-p}\int\Bigg[\,\,P[C^{(9-p)}]\,+\,
{\cal F}\wedge P[C^{(7-p)}]\,\,\Bigg]\,\,.
\label{veinte}
\eeq

The worldvolume coordinates $\xi^{\alpha}$ ($\alpha=0,\cdots,8-p$) will be
taken as:
\beq
\xi^{\alpha}\,=\,(t,x^{p-1},x^{p},\theta^1,\cdots,\theta^{6-p})\,\,.
\label{vuno}
\eeq
As we will confirm soon, the set of coordinates (\ref{vuno}) is quite
convenient to study the kind of configurations we are interested in. These
configurations are embeddings of the D(8-p)-brane which, in our system of
coordinates, are described by functions of the type:
\beq
r\,=\,r(t)\,\,,
\,\,\,\,\,\,\,\,\,\,\,\,\,
\rho\,=\,\rho(t)\,\,,
\,\,\,\,\,\,\,\,\,\,\,\,\,
\phi\,=\,\phi(t)\,\,.
\label{vdos}
\eeq
Let us now evaluate the action for the ansatz (\ref{vdos}). We shall begin by
studying the Wess-Zumino term. In this term only the components of $C^{(7-p)}$
and $C^{(9-p)}$ written in eq. (\ref{doce}) contribute. Actually, it is easy
to see that the pullback of the RR potential $C^{(7-p)}$ is coupled to the 
$x^{p-1}x^{p}$ component of ${\cal F}$. Assuming that ${\cal
F}_{x^{p-1},x^{p}}$ is independent of the angles $\theta^1\cdots\theta^{6-p}$,
one gets that $S_{WZ}$ can be written as:
\beq
S_{WZ}\,=\,T_{8-p}\,\Omega_{6-p}\,R^{7-p}\,\cos\varphi\,
\int dt\,dx^{p-1}\,dx^{p}\,
\rho^{7-p}\,(-1)^{p+1}\,\dot\phi\,
\Big[\,{\cal F}_{x^{p-1},x^{p}}\,+\,h_pf_p^{-1}\tan\varphi\,\Big]\,\,,
\label{vtres}
\eeq
where $\dot\phi=d\phi/dt$ and $\Omega_{6-p}$ is the volume of the $S^{6-p}$
sphere, given by:
\beq
\Omega_{6-p}\,=\,{2\pi^{{7-p\over 2}}\over 
\Gamma\Bigl(\,{7-p\over 2}\Bigr)}\,\,.
\label{vcuatro}
\eeq
Notice that for the ansatz (\ref{vdos}) the scalars $r$, $\rho$ and $\phi$ do
not depend on the coordinates $x^{p-1}$ and $x^{p}$ and, therefore, 
$P[B]$ has only non-zero components along the directions
$x^{p-1}x^{p}$. By using the actual value of the $B$ field for the 
(D(p-2), Dp) background in the definition of ${\cal F}$ (eq. (\ref{dnueve})),
one immediately gets that the term inside the square brackets on the 
right-hand side of eq. (\ref{vtres}) is:
\beq
{\cal F}_{x^{p-1},x^{p}}\,+\,h_pf_p^{-1}\tan\varphi\,=\,F_{x^{p-1},x^{p}}\,\,.
\label{vcinco}
\eeq
In what follows we will assume that the only non-zero component of the
worldvolume gauge field is $F_{x^{p-1},x^{p}}$ and we will denote from now on 
${\cal F}_{x^{p-1},x^{p}}$ and $F_{x^{p-1},x^{p}}$ simply by ${\cal F}$ and $F$
respectively. Then, the total action can be written as:
\beq
S\,=\,\int\,dt\,dx^{p-1}\,dx^p\,{\cal L}\,\,,
\label{vseis}
\eeq
where the lagrangian density ${\cal L}$ is given by:
\bear
{\cal L}&=&T_{8-p}\,\Omega_{6-p}\,R^{7-p}\,\times\rc\rc
&&\times\Bigg[\,-\rho^{6-p}\lambda_1\,
\sqrt{r^{-2}\,f_p^{-1}\,\,-\,r^{-2}\dot r^2\,-\,
{\dot\rho^2\over 1-\rho^2}\,-\,
(1-\rho^2)\,\dot\phi^2}\,+\,\lambda_2\,(-1)^{p+1}\rho^{7-p}\,\dot\phi
\,\,\Bigg]\,\,.\rc\rc
\label{vsiete}
\eear
In eq. (\ref{vsiete}) we have introduced the functions $\lambda_1$ and 
$\lambda_2$, which are defined as:
\beq
\lambda_1\,=\,\sqrt{h_pf_p^{-1}\,+\,{\cal F}^2\,h_p^{-1}}\,\,,
\,\,\,\,\,\,\,\,\,\,\,\,\,\,\,\,\,\,\,\,\,\,\,\,\,\,
\lambda_2\,=\,F\cos\varphi\,\,.
\label{vocho}
\eeq
Before starting the analysis of the lagrangian density (\ref{vsiete}), let us
discuss how the brane probe is extended along the $x^{p-1}x^{p}$ directions.
The main motivation to extend the probe along these directions is to allow the
worldvolume of the brane to capture the flux of the $B$ field of the
background. It is thus natural to characterize the spreading of the
D(8-p)-brane in the $x^{p-1}x^{p}$ plane by means of the flux of the $F$ field.
Accordingly we will extend our brane along the $x^{p-1}\,x^p$ directions in such
a way that there are $N'$ units of worldvolume flux. \ie :
\beq
\int\,\, dx^{p-1}\,dx^p\,F\,=\,{2\pi\over T_f}\,\,N'\,\,,
\label{vnueve}
\eeq
with $T_f=(2\pi\alpha')^{-1}$ being the fundamental string tension. Clearly,
given $F$ (which will be determined below), eq. (\ref{vnueve}) gives the volume
occupied by the brane probe in the noncommutative plane in terms of the flux
number $N'$.

Let us now resume our study of the dynamics of the system by performing a
canonical hamiltonian analysis of the lagrangian density  (\ref{vsiete}). First
of all, for simplicity, let us absorb the sign $(-1)^{p+1}$ sign of the
Wess-Zumino term of ${\cal L}$ by redefining
$\dot\phi$ if necessary. The density of momenta associated to ${\cal L}$  are:
\bear
{\cal P}_r&=&{\partial {\cal L}\over \partial \dot r}\,\equiv\,
T_{8-p}\,\Omega_{6-p}\,R^{7-p}\,\lambda_1\,\pi_r\,\,,\rc\rc
{\cal P}_{\rho}&=&{\partial {\cal L}\over \partial \dot \rho}\,\equiv\,
T_{8-p}\,\Omega_{6-p}\,R^{7-p}\,\lambda_1\,\pi_{\rho}\,\,,\rc\rc
{\cal P}_{\phi}&=&{\partial {\cal L}\over \partial \dot \phi}\,\equiv\,
T_{8-p}\,\Omega_{6-p}\,R^{7-p}\,\lambda_1\,\pi_{\phi}\,\,,
\label{treinta}
\eear
where we have defined the reduced momenta $\pi_r$, $\pi_{\rho}$ and 
$\pi_{\phi}$. By using the explicit value of  of ${\cal L}$,
given in eq. (\ref{vsiete}), we get:
\bear
 \pi_r&=&{\rho^{6-p}\over r^2}\,\,{\dot r\over
\sqrt{r^{-2}\,f_p^{-1}\,\,-\,r^{-2}\dot r^2\,-\,
{\dot\rho^2\over 1-\rho^2}\,-\,(1-\rho^2)\,\dot\phi^2}}\,\,,\rc\rc
\pi_{\rho}&=&{\rho^{6-p}\over 1- \rho^2}\,\,{\dot \rho\over
\sqrt{r^{-2}\,f_p^{-1}\,\,-\,r^{-2}\dot r^2\,-\,
{\dot\rho^2\over 1-\rho^2}\,-\,(1-\rho^2)\,\dot\phi^2}}\,\,,\rc\rc
 \pi_{\phi}&=&(1-\rho^2)\rho^{6-p}\,\,{\dot \phi\over
\sqrt{r^{-2}\,f_p^{-1}\,\,-\,r^{-2}\dot r^2\,-\,
{\dot\rho^2\over 1-\rho^2}\,-\,(1-\rho^2)\,\dot\phi^2}}
\,\,+\,\,\Lambda\,\rho^{7-p}\,\,,\rc
\label{tuno}
\eear
where, in the third of these expressions we have introduced the quantity
$\Lambda$, defined as:
\beq
\Lambda\,=\,{\lambda_2\over \lambda_1}\,\,.
\label{tdos}
\eeq
The hamiltonian density of the system can be obtained in the standard way,
namely:
\beq
{\cal H}\,=\,\dot r\,{\cal P}_r\,+\,\dot\rho\,{\cal P}_{\rho}\,
+\,\dot\phi{\cal P}_{\phi}\,-\,{\cal L}\,\equiv\,
\,T_{8-p}\,\Omega_{6-p}\,R^{7-p}\,\lambda_1\,h\,\,,
\label{ttres}
\eeq
where, in analogy with what we have done for the momenta, we have defined the
reduced quantity $h$. From eqs. (\ref{vsiete}) and (\ref{tuno}), one gets after
a simple calculation that $h$ is given by:
\beq
h\,=\,r^{-1}\,f_p^{-{1\over 2}}\,\Bigg[\,r^2\,\pi_r^2\,+\,\rho^{2(6-p)}\,+\,
(1-\rho^2)\,\pi_{\rho}^2\,+\,
{\Big(\pi_{\phi}-\Lambda\rho^{7-p}\Big)^2\over 1-\rho^2}
\,\,\Bigg]^{{1\over 2}}\,\,.
\label{tcuatro}
\eeq

\medskip
\subsection{Fixed  size configurations}
\medskip

We would like to obtain solutions of the equations of motion derived from the
hamiltonian (\ref{tcuatro}) which correspond to a brane of fixed size. It
follows from eq. (\ref{ocho}) that the coordinate $\rho$ plays the role of the
size of the system on the $S^{6-p}$ sphere. For this reason it is interesting
to look at solutions of the equations of motion for which $\rho$ is constant.
This same problem was analyzed in ref. \cite{DTV} for the case of brane probes
moving in the near-horizon Dp-brane background (see also refs. \cite{GST, GMT,
HHI, DJM}).  In ref. \cite{DTV} it was found that the $\rho$-dependent terms in
the hamiltonian can be arranged to produce a sum of squares from which the
$\rho={\rm constant}$ solutions can be found by inspection. By comparing the
right-hand side of eq. (\ref{tcuatro}) with the corresponding expression in ref.
\cite{DTV}, one immediately realizes that the same kind of arrangement can be
done in our case if the condition:
\beq
\Lambda\,=\,1
\label{tcinco}\,\,,
\eeq
is satisfied. Indeed, if eq. (\ref{tcinco}) holds, one can rewrite $h$ as:
\beq
h\,=\,r^{-1}\,f_p^{-{1\over 2}}\,\Bigg[\,\pi_{\phi}^2\,+\,r^2\,\pi_r^2\,+\,
(1-\rho^2)\,\pi_{\rho}^2\,+\,
{\Big(\pi_{\phi}\rho-\rho^{6-p}\Big)^2\over 1-\rho^2}
\,\,\Bigg]^{{1\over 2}}\,\,.
\label{tseis}
\eeq
The analysis of the hamiltonian (\ref{tseis}) can be performed along the same
lines as in ref. \cite{DTV} (see below). Before carrying out this study, let us
explore the content of the condition (\ref{tcinco}). By recalling the definition
of
$\Lambda$ (eq. (\ref{tdos})), we realize that eq.  (\ref{tcinco}) is equivalent
to require that $\lambda_1=\lambda_2$. Moreover, by using eq. (\ref{vcinco}) in
the definition of $\lambda_1$ (eq. (\ref{vocho})), one obtains:
\beq
\lambda_1^2\,=\,\cos^2\varphi\,F^2\,+\,f_p^{-1}\,\Big(\,F\sin\varphi\,-\,
{1\over \cos\varphi}\Big)^{2}\,\,.
\label{tsiete}
\eeq
By comparing the right-hand side of eq. (\ref{tsiete}) with the definition of 
$\lambda_2$ (eq. (\ref{vocho})), one concludes immediately that the condition
$\Lambda=1$ is equivalent to have the following constant value of the
worldvolume gauge field $F$:
\beq
F\,=\,{1\over \sin\varphi\cos\varphi}\,=\,2\csc (2\varphi)\,\,.
\label{tocho}
\eeq
By substituting this value of $F$ on the right-hand side of eq.
(\ref{tsiete}) one gets that $\lambda_1$ is also constant and given by:
\beq
\lambda_1\,=\,{1\over \sin\varphi}\,\,.
\label{tnueve}
\eeq
In the next section we will show that for the value of $F$ displayed in eq.
(\ref{tocho}) the brane probe breaks the  supersymmetry of the background
exactly in the same way as a massless particle. In this section this value of
the worldvolume gauge field should be considered as an ansatz which allows us
to find a class of fixed size solutions of the equations of
motion of the brane which are particularly interesting. In the remaining
of this section we will assume that $F$ is given by (\ref{tocho}).

It is now straightforward to find the configurations of the system with
constant $\rho$. From the second expression in eq. (\ref{tuno}) one concludes
that one must have:
\beq
\pi_{\rho}\,=\,0\,\,.
\label{cuarenta}
\eeq
Then, the hamiltonian equation of motion for $\pi_{\rho}$, \ie\ 
$\dot\pi_{\rho}\,=-\,\partial h / \partial \rho$, implies that the last term on
the right-hand side of eq. (\ref{tseis}) must vanish. For $p<6$ this happens
either for:
\beq
\rho=0\,\,,
\label{cuno}
\eeq
or else when $\pi_{\phi}$ is given by:
\beq
\pi_{\phi}\,=\,\rho^{5-p}\,\,.
\label{cdos}
\eeq 
(when $p=6$ only eq. (\ref{cdos}) gives rise to a constant $\rho$
configuration). Notice that, as $h$ does not depend on $\phi$, the
momentum $\pi_{\phi}$ is a constant of motion and, thus,  for $p\not= 5$, eq.
(\ref{cdos}) only makes sense when $\rho$ is constant. Moreover, when 
$p\not= 5$, eq. (\ref{cdos}) determines $\rho$ in terms of $\pi_{\phi}$. On the
contrary, when $p= 5$ eq. (\ref{cdos})  fixes $\pi_{\phi}\,=\,1$, independently
of the value of $\rho$. 

It is not difficult to translate the constant $\rho$ condition into 
a relation involving the time derivatives of $\phi$ and $r$. Let us, first of
all, invert the relation (\ref{tuno}) between $\pi_{\phi}$ and $\dot\phi$. By
taking $\Lambda=1$ as in eq. (\ref{tcinco}) one easily finds after a simple
calculation that:
\beq
\dot\phi\,=\,
{\pi_{\phi}-\rho^{7-p}\over 1-\rho^2}\,\,
{\Bigg[\,r^{-2}\big(f_p^{-1}\,-\,\dot r^2\,\big)\,-\,
{\dot\rho^2\over 1-\rho^2}\,\Bigg]^{{1\over 2}}\over
\Bigg[\,\pi_{\phi}^2\,+\,
{\big(\,\pi_{\phi}\rho\,-\,\rho^{6-p}\,\big)^2
\over 1-\rho^2}\,\Bigg]^{{1\over 2}}}\,\,.
\label{cdosextra}
\eeq
By taking $\dot\rho=0$ on  eq. (\ref{cdosextra}), and imposing one of the two
conditions (\ref{cuno}) or (\ref{cdos}), one gets that in both cases $\dot\phi$
and $\dot r$ satisfy the following relation:
\beq
f_p\,\big[\,r^2\,\dot\phi^2\,+\,\dot r^2\,]\,=\,1\,\,.
\label{ctres}
\eeq
For the configurations we are considering the last two terms of the reduced
hamiltonian $h$ of eq. (\ref{tseis}) vanish and, then, our configurations
certainly minimize the energy. These configurations are characterized by eq.  
 (\ref{tocho}), which fixes the value of
the worldvolume gauge field, and  (\ref{ctres}), which is a consequence of
the vanishing of the last two terms of the hamiltonian. Remarkably, eq.
(\ref{ctres}) is the condition satisfied by a particle moving in the $(r,\phi)$
plane at $\rho=0$ along a null trajectory, \ie\ with
$ds^2=0$, in the metric (\ref{uno}). Thus, our brane probe configurations have
the  characteristics of a massless particle : the so-called giant graviton.
Notice that the point $\rho=0$ can be considered as the ``center of mass" of
the expanded brane. In
order to confirm this picture let  us introduce the two-component vector 
${\bf v}$, defined as:
\beq
{\bf v}\,=\,(v^{\underline{r}}, v^{\underline{\phi}})\,\equiv\,
f_p^{{1\over 2}}\,(\dot r\,,\,r\dot\phi)\,\,,
\label{ccuatro}
\eeq
which is nothing but the velocity of the particle in the $(r,\phi)$ plane. From
eq. (\ref{ctres}) one clearly has:
\beq
(v^{\underline{r}})^2\,+\,(v^{\underline{\phi}})^2\,=\,1\,\,,
\label{ccinco}
\eeq
which simply states that the center of mass of the giant graviton moves
at the speed of light.  The corresponding value of the momentum density ${\cal
P}_{\phi}$ can be  straightforwardly obtained from eqs. (\ref{cdos}) and
(\ref{treinta}). Indeed, by using eqs.  (\ref{docho}), (\ref{vcuatro}),
(\ref{tres}) and (\ref{tnueve}), one gets:
\beq
{\cal P}_{\phi}\,=\,{T_f\over 2\pi}\,F\,N\,\rho^{5-p}\,\,,
\label{cseis}
\eeq
where $F$ is given in eq. (\ref{tocho}). 
The momenta $p_{\phi}$ and $p_r$ can be obtained from the densities
${\cal P}_{\phi}$ and ${\cal P}_{r}$ by integrating them in the $x^{p-1}x^p$
plane:
\beq
p_\phi\,=\,\int dx^{p-1}\,dx^p\,\,{\cal P}_{\phi}\,\,,
\,\,\,\,\,\,\,\,\,\,\,\,\,\,\,\,\,\,\,\,\,\,
p_r\,=\,\int dx^{p-1}\,dx^p\,\,{\cal P}_{r}\,\,.
\label{csiete}
\eeq
By using the value of the momentum density ${\cal P}_{\phi}$ given in eq. 
(\ref{cseis}) and the flux condition (\ref{vnueve}), one gets the following
value  of $p_\phi$ for a giant graviton:
\beq
p_\phi\,=\,N\,N'\,\rho^{5-p}\,\,.
\label{cocho}
\eeq
It follows from eq. (\ref{cocho}) that, when $p<5$, the size $\rho$ of the
wrapped brane increases with the momentum $p_\phi$. Moreover,  it is interesting
to point out that, as $0\le \rho\le 1$, when
$p<5$ the momentum $p_\phi$ has a maximum value $p_\phi^{max}$ given by:
\beq
p_\phi^{max}\,=\,N\,N'\,\,,
\label{cnueve}
\eeq
which is reached when $\rho=1$. The existence of such a maximum for $p_\phi$ is
a manifestation of the so-called stringy exclusion principle. Notice that when
$p=5$ the momentum $p_\phi$ is independent of $\rho$, whereas for $p=6$ the
value of $p_\phi$ written in eq. (\ref{cnueve}) is in fact a minimum. 

Let us now study the energy of the giant graviton solution. First of all, we
define the metric elements ${\cal G}_{MN}$ as:
\beq
{\cal G}_{MN}\,=\,{G_{MN}}_{{\,\big |}_{\rho=0}}\,\,,
\label{cincuenta}
\eeq
where the $G_{MN}$'s correspond to the metric displayed in eqs. (\ref{uno}) and
(\ref{ocho}). The hamiltonian for the giant graviton configurations, which we
will denote by $H_{GG}$, can be easily obtained from eq. (\ref{tseis}). In
terms of the $\rho=0$ metric it can be written as:
\beq
 H_{GG}\,=\,\sqrt{-{\cal G}_{tt}}\,\Bigg[\,
 {p_{\phi}^2\over {\cal G}_{\phi\phi}}\,+\,
{ p_{r}^2\over {\cal G}_{rr}}\,\,\Bigg]^{{1\over 2}}\,\,,
\label{ciuno}
\eeq
which, according to our expectations, is exactly the hamiltonian of a massless
particle which moves in the metric ${\cal G}_{MN}$ along a trajectory contained
in the $(r,\phi)$ plane. Interestingly, one can use the hamiltonian 
(\ref{ciuno}) to find and solve the equations of motion for the giant graviton.
Let us, first of all, rewrite $H_{GG}$ as:
\beq
 H_{GG}\,=\,R^{{p-7\over 2}}\,\,
\Bigg[\,r^{7-p}\, p_{r}^2\,+\,r^{5-p}\, p_{\phi}^2
\,\Bigg]^{{1\over 2}}\,\,.
\label{cidos}
\eeq
We will study the equations of motion of this system by using the conservation
of energy. In this method we first put $H_{GG}\,=\,E$, for constant $E$, and
then we use the relation between $p_{r}$ and $\dot r$, namely:
\beq
p_{r}\,=\,{R^{7-p}\over r^{7-p}}\,\,E\,\,\dot r\,\,.
\label{citres}
\eeq
By substituting (\ref{citres}) in the condition $H_{GG}\,=\,E$, we get:
\beq
\dot r^2\,+\,{r^{7-p}\over R^{7-p}}\,\Bigg[\,
{p_{\phi}^2\over E^2 R^{7-p}}\,r^{5-p}\,-\,1\,\Bigg]\,=\,0\,\,.
\label{cicuatro}
\eeq
Eq. (\ref{cicuatro}) determines the range of values that $r$ can take. Indeed,
by consistency of eq. (\ref{cicuatro}), the second term in this equation must
be negative or null. The points for which this term vanishes are the turning
points of the system. For $p<5$ these points are $r=0$ and $r_{*}$, with 
$r_{*}$ being:
\beq
(r_{*})^{5-p}\,=\,{E^2\over p_{\phi}^2}\,R^{7-p}\,\,.
\label{cicinco}
\eeq
In this $p<5$ case, $r$ can take values in the range $0\le r\le r_{*}$. If 
$p=5$ only the $r=0$ turning point exists and $r$ is unrestricted, \ie\ $r$
can take any non-negative value. Finally if $p=6$ the $r=0$ turning point is
missing and $r\ge r_{*}$, where $r_{*}$ is the value given in eq. 
(\ref{cicinco}) for $p=6$. 

It is not difficult to find the explicit dependence of $r$ on $t$. Let us
consider first the $p\not= 5$ case. From eq. (\ref{cicuatro}) it follows that
$t$ as a function of $r$ is given by the following indefinite integral:
\beq
t\,-\,t_{*}\,=\,R^{{7-p\over 2}}\,\int
{dr\over r^{{7-p\over 2}}\,
\sqrt{1\,-\,\Big(\,{r\over r_{*}}\,\Big)^{5-p}}}\,\,,
\,\,\,\,\,\,\,\,\,\,\,\,\,\,\,\,\,\,\,\,\,\,\,\,\,\,\,\,\,\,\,\,
(\,p\not= 5\,)\,\,,
\label{ciseis}
\eeq
where $t_{*}$ is a constant of integration. The integral (\ref{ciseis}) can be
easily performed by means of the following trigonometric change of variables:
\beq
\Big(\,{r\over r_{*}}\,\Big)^{5-p}\,=\,\cos^2\theta\,\,,
\,\,\,\,\,\,\,\,\,\,\,\,\,\,\,\,\,\,\,\,\,\,\,\,\,\,\,\,\,\,\,\,
(\,p\not= 5\,)\,\,.
\label{cisiete}
\eeq
The result of the integration  is:
\beq
\Big(\,{r_{*}\over r}\,\Big)^{5-p}\,=\,1\,+\,
(5-p)^2\,(r_{*})^{5-p}\,R^{p-7}\,
\Big(\,{t\,-\,t_{*}\over 2}\,\Big)^{2}\,\,,
\,\,\,\,\,\,\,\,\,\,\,\,\,\,\,\,\,\,\,\,\,\,\,\,\,\,\,\,\,\,\,\,
(\,p\not= 5\,)\,\,.
\label{ciocho}
\eeq 
It follows from eq. (\ref{ciocho}) that $t_{*}$ is precisely the value of $t$ 
at which $r=r_{*}$. Moreover if $p<5$ the coordinate $r\rightarrow 0$ as
$t\,-\,t_{*}\rightarrow \pm \infty$, \ie\ the giant graviton always falls
asymptotically into the black hole. On the contrary, for $p=6$ the coordinate
$r$ diverges asymptotically and, thus, in this case the particle always escapes
away from the $r=0$ point.

The  $p=5$ case needs a special treatment since, in this case, eq.
(\ref{ciseis}) is not valid any more. One can, however, easily integrate eq. 
(\ref{cicuatro}) with the result:
\beq
r\,=\,r_0\,e^{\pm{t\over R}\,\,\sqrt{1\,-\,{p_{\phi}^2\over E^2 R^2}}\,\,},
\,\,\,\,\,\,\,\,\,\,\,\,\,\,\,\,\,\,\,\,\,\,\,\,\,\,\,\,\,\,\,\,
(\,p= 5\,)\,\,.
\label{cinueve}
\eeq
It follows from (\ref{cinueve}) that, in this case, the solution connects  
asymptotically the points $r=0$ and $r=\infty$. 

In order to complete the integration of the equations of motion one has to
determine $\phi$ as a function of $t$. This can be easily achieved by
substituting $r(t)$ from eq. (\ref{ciocho}) or (\ref{cinueve}) into the
condition (\ref{ctres})  to get $\dot\phi$, followed by an integration over
$t$. The result for $p\not= 5$ is:
\beq
\tan\Bigg[\,{5-p\over 2}\,(\phi\,-\,\phi_{*})\,\Bigg]\,=\,
{5-p\over 2}\,\Bigg(\,{r_{*}\over R}\,\Bigg)^{{5-p\over 2}}\,\,\,
{t-t_{*}\over R}\,\,,
\,\,\,\,\,\,\,\,\,\,\,\,\,\,\,\,\,\,\,\,\,\,\,\,\,\,\,\,\,\,\,\,
(\,p\not= 5\,)\,\,,
\label{sesenta}
\eeq
whereas for $p= 5$ one gets:
\beq
\phi\,=\,\phi_0\,+\,{p_{\phi}\over E\,R^2}\,\,t\,\,,
\,\,\,\,\,\,\,\,\,\,\,\,\,\,\,\,\,\,\,\,\,\,\,\,\,\,\,\,\,\,\,\,
(\,p= 5\,)\,\,.
\label{suno}
\eeq
It is also interesting to find the value of the velocity ${\bf v}$ along the
trajectory. Actually, a simple calculation shows that, for $p\not= 5$, it depends
on the coordinate $r$ as:
\beq
v^{\underline{r}}\,=\,-\Big[\,1\,-\,\Big(\,{r\over r_*}\,\Big)^{5-p}
\,\Big]^{{1\over 2}}\,\,,
\,\,\,\,\,\,\,\,\,\,\,\,\,\,\,\,\,\,\,\,\,\,\,\,\,\,\,
v^{\underline{\phi}}\,=\,\Big(\,{r\over r_*}\,\Big)^{{5-p\over 2}}\,\,,
\,\,\,\,\,\,\,\,\,\,\,(p\not= 5)\,\,,
\label{sunoextrauno}
\eeq
where, on the right-hand side, it should be understood that $r$ is the function
of $t$ given in eq. (\ref{ciocho}) for $t\ge t_*$.  Curiously, when $p=5$ the
vector
${\bf v}$ is constant and has the following components:
\beq
v^{\underline{r}}\,=\,\pm\Big[\,1\,-\,{p_{\phi}^2\over E^2\,R^2}
\,\Big]^{{1\over 2}}\,\,,
\,\,\,\,\,\,\,\,\,\,\,\,\,\,\,\,\,\,\,\,\,\,\,\,\,\,\,
v^{\underline{\phi}}\,=\,{p_{\phi}\over E\,R}\,\,,
\,\,\,\,\,\,\,\,\,\,\,(p= 5)\,\,.
\label{sunoextrados}
\eeq
To finish this section, 
let us now discuss the extensivity of the brane probe on the noncommutative
plane $x^{p-1}\,x^p$. As argued above, for fixed worldvolume flux $N'$, this
extensivity depends on the value of the gauge field $F$. Actually, by using in
eq. (\ref{vnueve}) the  value of $F$ given in eq. (\ref{tocho}), one gets
that the volume occupied by the brane probe along the $x^{p-1}\,x^p$ plane is:
\beq
\int\,\, dx^{p-1}\,dx^p\,=\,{\pi N'\over T_f}\,\,\sin(2\varphi)\,\,.
\label{sdos}
\eeq
This volume clearly goes to zero as $\varphi\rightarrow 0$ for fixed $N'$. This
means that if we switch off the $B$ field in the background, the D(8-p)-brane
probe is effectively converted into a D(6-p)-brane, in agreement with the
results of refs. \cite{GST}-\cite{DTV}. Actually, one can check that in this
limit our results agree with those corresponding to the Dp-brane background if we
replace everywhere $N$ by $NN'$.

\setcounter{equation}{0}
\section{Supersymmetry}
\medskip
The objective of this section is to analyze the  supersymmetry behaviour of the 
brane configurations studied in section 2. Actually, we will verify that 
these configurations break the supersymmetry of the background
just as  a massless particle which moves precisely along the trajectories found
in section 2.

The number of supersymmetries preserved by a Dp-brane is the number of
independent solutions of the equation \cite{bbs}:
\beq
\Gamma_{\kappa}\,\epsilon\,=\,\epsilon\,\,,
\label{stres}
\eeq
where $\epsilon$ is a Killing spinor of the background and $\Gamma_{\kappa}$ is
the so-called $\kappa$-symmetry matrix, which depends on the background and on
the type of brane. For a Dp-brane in a type IIB background $\Gamma_{\kappa}$ is
given by \cite{bbs}:
\bear
\Gamma_{\kappa}\,&=&\,{1\over 
\sqrt{-{\rm det}\,(\,g\,+\,{\cal F}\,)}}\,\,
\sum_{n=0}^{\infty}\,{1\over 2^n\,n!}\,\,
\gamma^{\mu_1\nu_1\cdots\mu_n\nu_n}\times\rc\rc\rc
&&\times\,{\cal F}_{\mu_1\nu_1}\,\cdots\,
{\cal F}_{\mu_n\nu_n}\,\,J^{(n)}\,\,,
\label{scuatro}
\eear
where $g$ is the induced metric on the brane worldvolume, ${\cal F}=F-P[B]$ and
$J^{(n)}$ is the following matrix:
\beq
J^{(n)}\,=\,(-1)^n\,\sigma_3^{{p-3\over
2}+n}\,(i\sigma_2)\,\otimes\,\Sigma_0\,\,,
\label{scinco}
\eeq
with $\Sigma_0$ being:
\beq
\Sigma_0\,=\,{1\over (p+1)!}\,\,\epsilon^{\mu_1\cdots\mu_{p+1}}\,
\gamma_{\mu_1\cdots\mu_{p+1}}\,\,.
\label{sseis}
\eeq
Recall that in the type IIB theory the spinor $\epsilon$ is actually composed
by two Majorana-Weyl spinors $\epsilon_1$ and $\epsilon_2$ of well defined
ten-dimensional chirality, which can be arranged as a two-component vector in
the form:
\beq
\epsilon\,=\,\pmatrix{\epsilon_1\cr\epsilon_2}\,\,.
\label{sseisextra}
\eeq
The Pauli matrices appearing in the expression of $J^{(n)}$ act on this
two-dimensional vector. Moreover, in  eqs. (\ref{scuatro}) and (\ref{scinco})
$\gamma_{\mu_1\mu_2\cdots}$ are
antisymmetrized products of the induced world-volume Dirac matrices
which, in terms of the ten-dimensional constant gamma matrices 
$\Gamma_{\underline{M}}$, are given by:
\beq
\gamma_{\mu}\,=\,
\partial_{\mu}\,X^M\,E_{M}^{\underline{M}}\,
\Gamma_{\underline{M}}\,\,,
\label{ssiete}
\eeq
where $E_{M}^{\underline{M}}$ is the ten-dimensional vielbein. There exist
similar expressions for the type IIA theory. Actually, for the sake of
simplicity, we will restrict ourselves to the analysis of the $(D1,D3)$
background, although one can find similar results for the general case by
making the appropriate modifications in our equations.

\medskip
\subsection{Killing spinors of the $\bf {(D1,D3)}$ background}
\medskip

In order to solve eq. (\ref{stres}) it is clear that we must first determine
the Killing spinors of the background, which is equivalent to characterize the
supersymmetry preserved by our supergravity solution.  We
are considering bosonic  backgrounds, which are supersymmetric iff the
supersymmetry variation of the fermionic supergravity fields vanishes.
Actually, this only occurs for some class of transformation parameters, which
are precisely the Killing spinors we are interested in. 

The analysis of the supersymmetry preserved by the (D1,D3) background by using
the transformation rules of the type IIB supergravity is performed in appendix
A. Here we will characterize the Killing spinors of this background by means
of an alternative and more simplified method \cite{Kim}, which makes use of the
$\kappa$-symmetry matrix  (\ref{scuatro}). Let us place a test D3-brane parallel
to the background and choose its worldvolume coordinates as
$(\xi^{0}, \xi^{i})\,=\,(t,x^i)$ for $i\,=1,2,3$. The equations
which determine the embedding of the parallel D3-brane are :
\beq
X^{0}\,=\,t\,=\xi^0\,,
\,\,\,\,\,\,\,\,\,\,\,\,\,\,\,\,\,\,\,\,\,\,\,
X^{i}\,=\,x^i\,=\xi^i\,\,,\,\,\,\,\,\,\,\,\,\,\,\,(i\,=1,2,3)\,\,,
\label{socho}
\eeq
with the other spacetime coordinates being independent of the 
$(\xi^{0}, \xi^{i})$. We will study the supersymmetry preserved by this
test brane when the worldvolume gauge field $F$ is zero, \ie\ when 
${\cal F}\,=\,-P[B]$. This means that the only non-zero component of ${\cal F}$
is:
\beq
{\cal F}_{x^2x^3}\,=\,-\tan\varphi\,f_3^{-1}\,h_3\,\,.
\label{snueve}
\eeq
The test brane configuration we are considering is the same as the one that
creates the background. Therefore, it is natural to expect that such a parallel
test brane preserves the same supersymmetries as the supergravity background.
Then, let us consider the $\kappa$-symmetry condition 
$\Gamma_k\epsilon\,=\,\epsilon$ for this case. The induced gamma matrices for
the embedding (\ref{socho}) are:
\bear
\gamma_{x^{\alpha}}\,&=&\,f_3^{-{1\over 4}}\,\Gamma_{\underline x^{\alpha}}\,\,,
\,\,\,\,\,\,\,\,\,\,\,\,\,\,\,\,\,\,\,\,\,\,\,\,\,\,\,\,
(\alpha\,=0,1)\,\,,\rc\rc
\gamma_{x^{i}}\,&=&\,f_3^{-{1\over 4}}\,h_3^{{1\over 2}}\,
\Gamma_{\underline x^{i}}\,\,,
\,\,\,\,\,\,\,\,\,\,\,\,\,\,\,\,\,\,\,\,\,\,\,(i\,=2,3)\,\,.
\label{setenta}
\eear
By using eq. (\ref{setenta}) in the expression of $\Gamma_k$ for $p=3$ (eqs. 
(\ref{scuatro})-(\ref{sseis})), one gets that the $\kappa$-symmetry matrix for
this case is:
\beq
\Gamma_k\,=\,h_3^{{1\over 2}}\,\Big[\,\cos\varphi\,(i\sigma_2)\,
\Gamma_{\underline{x^0\cdots x^3}}\,-\,
f_3^{-{1\over 2}}\,\sin\varphi\,\sigma_1\Gamma_{\underline{x^0x^1}}
\,\Big]\,\,.
\label{stuno}
\eeq
In eq. (\ref{stuno}), and in what follows, we have suppressed the tensor
product symbol. It is now straightforward to check that the condition
$\Gamma_k\epsilon\,=\,\epsilon$ can be written as:
\beq
(i\sigma_2)\,
\Gamma_{\underline{x^0\cdots x^3}}\,\epsilon\,=\,
e^{-\alpha\,\Gamma_{\underline{x^2x^3}}\sigma_3}\,\epsilon\,\,,
\label{stdos}
\eeq
where $\alpha$ is the function of $r$ given by:
\beq
\sin\alpha\,=\,f_3^{-{1\over 2}}\,h_3^{{1\over 2}}\sin\varphi\,\,,
\,\,\,\,\,\,\,\,\,\,\,\,\,\,\,\,\,\,\,\,\,\,
\cos\alpha\,=\,h_3^{{1\over 2}}\cos\varphi\,\,.
\label{sttres}
\eeq
In appendix A we will verify that the supersymmetry invariance of the dilatino
of type IIB supergravity in the $(D1,D3)$ background requires that the
supersymmetry parameter $\epsilon$ satisfies eq. (\ref{stdos}) with $\alpha$
given by eq. (\ref{sttres}), which is is a confirmation of the correctness of our
$\kappa$-symmetry argument. Moreover, in view of eq. (\ref{stdos}), the Killing
spinor $\epsilon$ can be written as:
\beq
\epsilon\,=\,e^{{\alpha\over 2}\,\Gamma_{\underline{x^{2}x^{3}}}\,
\sigma_3}\,\,\tilde\epsilon\,\,,
\label{stcuatro}
\eeq
where $\tilde\epsilon$ is a spinor which satisfies:
\beq
(i\sigma_2)\,
\Gamma_{\underline{x^0\cdots x^3}}\,\,\tilde\epsilon\,\,=\,\,
\tilde\epsilon\,\,.
\label{stcinco}
\eeq

In our study of the supersymmetry preserved by the giant graviton solution we
will need to know the dependence of $\epsilon$ on the coordinate $\rho$. Again,
this dependence can be extracted  by solving the corresponding supergravity
equations. Here, however, we will present a simpler argument which, as we will
check in appendix A, gives the right answer. The starting point of our argument
is to consider the supergravity solution in the asymptotic region 
$r\rightarrow\infty$ without making the near-horizon approximation
(\ref{quince}). In this case the harmonic function $f_p$ is given by eq. 
(\ref{dos}) and, when $r\rightarrow\infty$,  the metric is flat
and the different forms vanish. The supersymmetry preserved in this region is
just the one corresponding to covariantly constant spinors, \ie\ spinors
which satisfy  $D_M\epsilon\,=\,0$ in the flat asymptotic metric. 
Taking $M\,=\,\rho$, we get that:
\beq
\partial_{\rho}\epsilon\,=\,-{1\over 4}\,\omega_{\rho}^{\underline {MN}}\,
\Gamma_{\underline {MN}}\,\epsilon\,\,,
\label{stseis}
\eeq
where $\omega_{\rho}^{\underline {MN}}$ are the components of the spin
connection. As the only non-zero component of the
form $\omega_{\rho}^{\underline {MN}}$ of the spin connection in our coordinates
(\ref{uno}) and (\ref{ocho}) for the the asymptotic metric is:
\beq
\omega_{\rho}^{\underline {\rho r}}\,=\,{1\over \sqrt{1-\rho^2}}\,\,,
\label{stsiete}
\eeq
we obtain from eq. (\ref{stseis})
that the $\rho$ dependence of $\tilde\epsilon$ can be
parametrized as:
\beq
\tilde\epsilon\,=\,
e^{-{\beta\over 2}\,\,
\Gamma_{\underline{\rho r} }}\,\,\,\hat\epsilon\,\,,
\label{stocho}
\eeq
with $\beta$ being the following function of $\rho$:
\beq
\sin\beta\,=\,\rho\,\,,
\,\,\,\,\,\,\,\,\,\,\,\,\,\,\,\,\,\,\,\,\,\,
\cos\beta\,=\,\sqrt{1-\rho^2}\,\,,
\label{stnueve}
\eeq
and $\hat\epsilon$ satisfying the same equation as $\tilde\epsilon$, namely:
\beq
(i\sigma_2)\,
\Gamma_{\underline{x^0\cdots x^3}}\,\,\hat\epsilon\,\,=\,\,
\hat\epsilon\,\,.
\label{ochenta}
\eeq
The supergravity analysis of the $(D1,D3)$ background provides the explicit
dependence of $\hat\epsilon$ on the other coordinates (see eq.
(\ref{apadoce})). However, in our analysis of the $\kappa$-symmetry preserved by
the giant graviton solution we will only need to know that $\hat\epsilon$
satisfies eq. (\ref{ochenta}). Actually, we will rewrite this equation in a form
more convenient for our purposes. Let us, first of all, recall that all spinors
(and in particular 
$\hat\epsilon$) have fixed chirality and thus $\hat\epsilon$ satisfies:
\beq
\Gamma_{\underline{x^0\cdots x^3}}\,\Gamma_{\underline{\rho r}}\,
\Gamma_{\underline{\phi}}\,\Gamma_{*}\,\hat\epsilon\,=\,\hat\epsilon\,\,,
\label{ouno}
\eeq
where we have denoted:
\beq
\Gamma_{*}\,=\,\Gamma_{\underline{\theta^{1}\cdots \theta^{3}}}\,\,.
\label{odos}
\eeq
Taking eq. (\ref{ouno}) into account, eq. (\ref{ochenta}) can be written as:
\beq
\Gamma_{\underline{\rho r}}\,\hat\epsilon\,=\,
\Upsilon\,\hat\epsilon\,\,,
\label{otres}
\eeq
where $\Upsilon$ is the following matrix:
\beq
\Upsilon\,=\,(i\sigma_2)\,\,\Gamma_{\underline{\phi}}\,\Gamma_{*}\,\,.
\label{ocuatro}
\eeq
Using eq. (\ref{otres}) in eq. (\ref{stocho}), we can  reexpress
$\tilde\epsilon$ as:
\beq
\tilde\epsilon\,=\,
e^{-{\beta\over 2}\,\,\Upsilon}\,\,
\,\,\,\hat\epsilon\,\,,
\label{ocinco}
\eeq
which is the parametrization of $\tilde\epsilon$ which we will use in our
analysis of the $\kappa$-symmetry for the giant graviton.

\medskip
\subsection{$\kappa$-symmetry of the probe}
\medskip

We are now ready to determine the supersymmetry preserved by the giant graviton
in the $(D1,D3)$ background\footnote{Similar methods have been applied in
refs. \cite{Imamura, kappa} to study the supersymmetry of the baryon vertex.}. 
According to the formalism developed in section 2,
we have to consider a D5-brane probe extended along the directions 
$(x^2, x^3, \theta^1, \theta^2, \theta^3)$. The $\kappa$-symmetry matrix 
$\Gamma_{\kappa}$ for such a probe can be obtained from eq. (\ref{scuatro})
by taking $p=5$. For a D5-brane embedding of the type (\ref{vdos}) with
$\dot\rho=0$ the induced Dirac matrices are:
\bear
\gamma_{x^0}&=&\sqrt{-G_{tt}}\,\Gamma_{\underline{x^0}}\,+\,
\dot\phi\sqrt{G_{\phi\phi}}\,\Gamma_{\underline{\phi}}\,+\,
\dot r\sqrt{G_{rr}}\,\Gamma_{\underline{r}}\,\,,\rc\rc
\gamma_{x^{2}x^{3}}&=&f_3^{-{1\over 2}}\,h_3\,
\Gamma_{\underline{x^{2}x^{3}}}\,\,,\rc\rc
\gamma_{\theta^i}&=&f_3^{{1\over 4}}\,r\,\rho \,e_{i}^{\,\underline i}
\,\Gamma_{\underline{\theta^{i}}}\,\,,
\label{oseis}
\eear
where $e_{j}^{\,\underline i}$ denotes the $S^{3}$ vielbein. We will assume, as
in section 2, that the brane probe has a worldvolume gauge field whose only
non-vanishing component is ${\cal F}_{x^2 x^3}$, which we simply will denote by
${\cal F}$. By substituting the values given in eq. (\ref{oseis}) in eq. 
(\ref{scuatro}) one readily verifies that the contribution of the gauge field
${\cal F}$ exponentiates and, as a consequence, the matrix 
$\Gamma_{\kappa}$ can be written as:
\bear
\Gamma_{\kappa}&=&{\,\,i\sigma_2\over
\sqrt{-G_{tt}\,-\,G_{\phi\phi}\dot\phi^2\,-\,G_{rr}\dot r^2}}\,\,\times\rc\rc
&&\times\Bigg[\,\sqrt{-G_{tt}}\,\Gamma_{\underline{x^0}}\,+\,
\dot\phi\sqrt{G_{\phi\phi}}\,\Gamma_{\underline{\phi}}\,+\,
\dot r\sqrt{G_{rr}}\,\Gamma_{\underline{r}}\,\Bigg]\,\Gamma_{*}\,\,
e^{-\eta\,\Gamma_{\underline{x^{2}x^{3}}}
\sigma_3}\,\,.
\label{osiete}
\eear
In eq. (\ref{osiete}) $G_{MN}$ denote the elements of the metric 
(\ref{uno}) for $p=3$, $\Gamma_{*}$ is the same as in eq. (\ref{odos}) and
$\eta$ is defined as:
\beq
\sin\eta\,=\,{f_3^{-{1\over 2}}\,h_3^{{1\over 2}}\over \lambda_1}\,\,,
\,\,\,\,\,\,\,\,\,\,\,\,\,\,\,\,\,\,\,\,\,\,
\cos\eta\,=\,{{\cal F}\,h_3^{-{1\over 2}}\over \lambda_1}\,\,,
\label{oocho}
\eeq
where $\lambda_1$ has been defined in eq. (\ref{vocho}).

Let us now compute the action of $\Gamma_{\kappa}$ on  spinor
$\epsilon$, which we will parametrize as the Killing spinors  of the $(D1,D3)$
background, namely (see eqs.  (\ref{stcuatro}) and  (\ref{ocinco})):
\beq
\epsilon\,=\,e^{{\alpha\over 2}\,\Gamma_{\underline{x^{2}x^{3}}}\,
\sigma_3}\,\,\tilde\epsilon\,=\,
e^{{\alpha\over 2}\,\Gamma_{\underline{x^{2}x^{3}}}\,
\sigma_3}\,\,
e^{-{\beta\over 2}\,\,\Upsilon}\,\,
\,\,\,\hat\epsilon\,\,,
\label{oochoextra}
\eeq
where $\alpha$ and $\beta$ are given in eqs. (\ref{sttres}) and (\ref{stnueve}),
$\Upsilon$ is the matrix written in (\ref{ocuatro}) and $\hat \epsilon$ is
independent of
$\rho$. By using this representation, one immediately gets:
\bear
\Gamma_{\kappa}\,\epsilon&=&e^{(\eta-{\alpha\over 2})\,\,
\Gamma_{\underline{x^{2}x^{3}}}\, \sigma_3}\,\,\,
{i\sigma_2\over
\sqrt{-G_{tt}\,-\,G_{\phi\phi}\dot\phi^2\,-\,G_{rr}\dot r^2}}\,\,\times\rc\rc
&&\times\Bigg[\,\sqrt{-G_{tt}}\,\Gamma_{\underline{x^0}}\,+\,
\dot\phi\sqrt{G_{\phi\phi}}\,\Gamma_{\underline{\phi}}\,+\,
\dot r\sqrt{G_{rr}}\,\Gamma_{\underline{r}}\,\Bigg]\,\Gamma_{*}\,\,
\tilde\epsilon\,\,.
\label{onueve}
\eear
Then, making use again of eq. (\ref{oochoextra}), one concludes that 
the equation $\Gamma_{\kappa}\,\epsilon=\epsilon$ is equivalent to
the following condition for $\tilde\epsilon$:
\bear
{i\sigma_2\over
\sqrt{-G_{tt}\,-\,G_{\phi\phi}\dot\phi^2\,-\,G_{rr}\dot r^2}}
\Bigg[\,\sqrt{-G_{tt}}\,\Gamma_{\underline{x^0}}+
\dot\phi\sqrt{G_{\phi\phi}}\,\Gamma_{\underline{\phi}}+
\dot r\sqrt{G_{rr}}\,\Gamma_{\underline{r}}\,\Bigg]\,\Gamma_{*}\,\,
\tilde\epsilon\,=
e^{(\alpha-\eta)\,\,
\Gamma_{\underline{x^{2}x^{3}}}\, \sigma_3}\,\tilde\epsilon\,\,.\rc
\label{noventa}
\eear
Notice that in eq. (\ref{noventa}) the matrix $\Gamma_{\underline{x^{2}x^{3}}}$
only appears on the right-hand side. Actually, if $\alpha=\eta$ this dependence 
on $\Gamma_{\underline{x^{2}x^{3}}}$ disappears. Moreover, by comparing the
definitions of $\alpha$ and $\eta$ (eqs. (\ref{sttres}) and (\ref{oocho}), 
respectively) one immediately realizes that $\alpha=\eta$ if and only if
$\lambda_1\,=\,1/\sin\varphi$. On the other hand, eq. (\ref{tsiete}) tells us
that this only happens when $F\,=\,2\csc(2\varphi)$, \ie\ precisely when the
worldvolume gauge field strength takes the same value as the one we have found in
our hamiltonian analysis of section 2 (see eq. (\ref{tocho})). Let us assume
that this is the case and let us try to find out what are the consequences of
this fact. Actually, we will also assume that our second condition 
(\ref{ctres}) holds. It is a simple exercise to verify that, when (\ref{ctres})
is satisfied, the denominator of the left-hand side of eq. (\ref{noventa})
takes the value:
\beq
\sqrt{-G_{tt}\,-\,G_{\phi\phi}\dot\phi^2\,-\,G_{rr}\dot r^2}\,=\,
\rho\,r\,f_3^{{1\over 4}}\,\dot\phi\,\,.
\label{nuno}
\eeq
Then, for a configuration satisfying eq. (\ref{tocho}) and (\ref{ctres}), 
the equation $\Gamma_{\kappa}\,\epsilon=\epsilon$ 
becomes:
\beq
\Bigg[\,\sqrt{-G_{tt}}\,\Gamma_{\underline{x^0\phi}}-
\dot\phi\sqrt{G_{\phi\phi}}+
\dot r\sqrt{G_{rr}}\,\Gamma_{\underline{r\phi}}\,\Bigg]\,\,
\tilde\epsilon\,=\,\rho\, r\, f_3^{{1\over 4}}\,\dot\phi\,\Upsilon
\,\,\tilde\epsilon\,\,,
\label{ndos}
\eeq
where $\Upsilon$ has been defined in eq. (\ref{ocuatro}). Let us now rewrite
eq. (\ref{ndos}) in terms of the $\rho=0$ spinor $\hat\epsilon$. By
substituting eq. (\ref{ocinco}) on both sides of eq. (\ref{ndos}), and using the
fact that $\Upsilon$ anticommutes with $\Gamma_{\underline{x^0\phi}}$ and 
$\Gamma_{\underline{r\phi}}$, one gets:
\beq
\Bigg[\,f_3^{-{1\over 4}}\,e^{\beta\,\Upsilon}\,
\Gamma_{\underline{x^0\phi}}-
\dot\phi\,r\,f_3^{{1\over 4}}\,\sqrt{1-\rho^2}+
\dot r f_3^{{1\over 4}}\,e^{\beta\,\Upsilon}\,
\Gamma_{\underline{r\phi}}\,\Bigg]\,\,
\hat\epsilon\,=\,\rho\, r\, f_3^{{1\over 4}}\,\dot\phi\,\Upsilon
\,\,\hat\epsilon\,\,.
\label{ntres}
\eeq
Let us consider now eq. (\ref{ntres}) for the particular case $\rho=0$. When
$\rho=0$ the right-hand side of eq. (\ref{ntres}) vanishes and, as 
$\beta=0$ is also zero in this case (see eq. (\ref{stnueve})), one has:
\beq
\Bigg[\,f_3^{-{1\over 4}}\,\Gamma_{\underline{x^0\phi}}
-\dot\phi\,r\,f_3^{{1\over 4}}+
\dot r f_3^{{1\over 4}}\Gamma_{\underline{r\phi}}\,\Bigg]\,\,
\hat\epsilon\,=\,0\,\,.
\label{ncuatro}
\eeq
For a general value of $\rho$ the $\kappa$-symmetry condition can be obtained
by substituting in (\ref{ntres}) $e^{\beta\,\Upsilon}$ by:
\beq
e^{\beta\,\Upsilon}\,=\,\sqrt{1-\rho^2}\,+\,\rho\,\Upsilon\,\,.
\label{ncinco}
\eeq
By so doing one obtains two types of terms, with and without $\Upsilon$, which,
amazingly, satisfy the equation independently for all $\rho$ as a consequence of
the $\rho=0$  condition (\ref{ncuatro}). Thus, eq. (\ref{ncuatro}) is
equivalent to the $\kappa$-symmetry condition 
$\Gamma_{\kappa}\epsilon=\epsilon$ and is the constraint we have to
impose to the Killing spinors of the background in order to define a
supersymmetry transformation preserved by our brane probe configurations. In
order to obtain a neat interpretation of (\ref{ncuatro}), let us define the
matrix $\Gamma_v$ as:
\beq
\Gamma_v\,\equiv\,v^{\underline{r}}\,\Gamma_{\underline{r}}\,+\,
v^{\underline{\phi}}\,\Gamma_{\underline{\phi}}\,\,,
\label{nseis}
\eeq
where $v^{\underline{r}}$ and $v^{\underline{\phi}}$ are the components of the
velocity vector ${\bf v}$ defined in eq. (\ref{ccuatro}). By using eq. 
(\ref{ccinco}), which is a consequence of (\ref{ctres}), one readily proves that
the matrix $\Gamma_v$ satisfies:
\beq
(\,\Gamma_v\,)^2\,=\,1\,\,.
\label{nsiete}
\eeq
Moreover, from the explicit expression of the components of ${\bf v}$ (see eq. 
(\ref{ccuatro})) it is straightforward to demonstrate that the $\kappa$-symmetry
condition (\ref{ncuatro}) can be written as:
\beq
\Big[\Gamma_{\underline{x^0}}\,+\,\Gamma_v\,\Big]\,\hat\epsilon\,=\,0\,\,.
\label{nocho}
\eeq
Taking into account eq. (\ref{nsiete}), 
one can  rewrite eq. (\ref{nocho}) in the form:
\beq
\Gamma_{\underline{x^0}}\,\Gamma_v\,\hat\epsilon\,\,=\,\,
\hat\epsilon\,\,.
\label{nochoextra}
\eeq
Moreover, recalling the relation (\ref{oochoextra}) between 
$\hat\epsilon$ and $\epsilon$, and taking into account that 
$\Gamma_{\underline{x^0}}\,\Gamma_v$ commutes with
$\Gamma_{\underline{x^{2}x^{3}}}$, eq. (\ref{nochoextra}) is equivalent to:
\beq
\Gamma_{\underline{x^0}}\,\Gamma_v\,\epsilon_{{\,\big |}_{\rho=0}}\,\,=\,\,
\epsilon_{{\,\big |}_{\rho=0}}\,\,.
\label{nnueve}
\eeq
Eq. (\ref{nnueve}) is the condition satisfied by the parameter of the
supersymmetry preserved by a massless particle
which moves  in the direction of the vector ${\bf v}$ at $\rho=0$, \ie\ by a
gravitational wave which propagates precisely along the trajectories found in
section 2. However, the background projector 
$(i\sigma_2)\, \Gamma_{\underline{x^0\cdots x^3}}$ and the one of the probe,
$\Gamma_{\underline{x^0}}\,\Gamma_v$ do not commute (actually, they
anticommute). This means that the conditions (\ref{ochenta}) and
(\ref{nochoextra}) cannot be imposed at the same time and, therefore, the probe
breaks completely the supersymmetry of the background. The most relevant
aspect of this result is that the supersymmetry breaking  produced by the probe
is just identical to the one corresponding to a massless particle which moves
in the direction of ${\bf v}$. This fact is a confirmation of our
interpretation of the giant graviton configurations as blown up gravitons.

Notice that we have found the supersymmetry projection for the
giant graviton configurations from the conditions (\ref{tocho}) and
(\ref{ctres}). Clearly, we could have done our reasoning in reverse order and,
instead, we could have imposed first that our brane probe breaks
supersymmetry as a massless particle at $\rho=0$. In this case we would arrive
at the same conditions as those obtained by studying the hamiltonian and,
actually, this would be an alternative way to derive them. 

\medskip
\setcounter{equation}{0}
\section{Summary and discussion}
\medskip

In this paper we have found configurations of a brane probe on the (D(p-2), Dp)
background which behave as a massless particle. We have checked this fact by
studying the motion of the brane and the way in which breaks
supersymmetry. These giant graviton configurations admit the interpretation of a
set of massless quanta polarized by the gauge fields of the background. Actually,
by recalling the arguments of refs. \cite{GST}-\cite{DTV}, one can argue that
there are two possible descriptions of this system, as a point-like particle or
as an expanded brane, which are valid for different ranges of the momenta and
cannot be simultaneously valid. 

In the cases studied in refs. \cite{GST}-\cite{DTV} the blow up of the gravitons
takes place on a (fuzzy) sphere. In our case the brane probe shares two
dimensions with the branes of the background and, thus, our gravitons are also
expanded along a noncommutative plane. We have parametrized the volume occupied
by the probe in the noncommutative plane by means of the flux of the worldvolume
gauge field. If this flux is fixed and finite, the angular momentum of the brane
is bounded for $p<5$ and one realizes the stringy exclusion principle. 

Let us finally comment on some possible extensions of our results. It is clear
that one should investigate the spectrum of small vibrations of the brane
around the giant graviton configuration, along the lines of refs.
\cite{DJM, KM}, in order to determine its stability. It would be also
very interesting to have a more explicit picture of the blow up of the gravitons
in the noncommutative plane in terms of the Myers dielectric efect. Another
topic which we would be interesting to examine is whether or not there exist
configurations similar to the ones studied here in M-theory. Notice that the
authors of the second paper in \cite{MR} found solutions of eleven dimensional
supergravity generated by a non-threshold  (M2,M5) bound state. The natural
probe to consider in this case is a M5-brane sharing  three common directions
with the background. We expect to report on these issues in future.

\medskip
\section{ Acknowledgments}
\medskip

We are grateful to J. Edelstein, A. Paredes, J. M. Sanchez de Santos and J.
Simon for discussions.  This work was
supported in part by DGICYT under grant PB96-0960,  by CICYT under
grant  AEN99-0589-CO2 and by Xunta de Galicia under  grant
PGIDT00-PXI-20609.

\vskip 2cm                                               
{\Large{\bf A. Supergravity analysis of the
(D1,D3) background}}                                 
\vskip .5cm                                              
\renewcommand{\theequation}{\rm{A}.\arabic{equation}}  
\setcounter{equation}{0}  

In this appendix we shall study the supersymmetry preserved by the $(D1,D3)$
solution of the type IIB supergravity equations. In order to perform this
analysis it is more convenient to work in the Einstein frame, in which the metric
$ds^2_E\,=\,e^{-\phi_D/2}\,ds^2$ takes the form:
\bear
ds^2_E&=&f_3^{-1/2}\,h_3^{-1/4}\,\Big[\,-(\,dx^0\,)^2\,+\,(\,dx^1\,)^2\,+
\,h_3\,\Big((\,dx^{2}\,)^2\,+\,(\,dx^{3}\,)^2\Big)\,\Big]\,+\rc\rc
&&+\,f_3^{1/2}\,h_3^{-1/4}\,dr^2\,+\,
f_3^{1/2}\,h_3^{-1/4}\,r^2\,\Big[\,{1\over 1-\rho^2}\,d\rho^2\,+\,
(1-\rho^2)d\phi^2\,+\,\rho^2d\Omega_3^2\,\Big]\,\,,\rc
\label{apauno}
\eear
where $f_3$ and $h_3$ are given in eq. (\ref{dos}) and, for simplicity, we have
taken the string coupling constant $g_s$ equal to one. 

The supersymmetry transformations of the type IIB supergravity fields have been
obtained in ref. \cite{SUSYIIB}. In this reference the author uses complex
spinors instead of working the real two-component spinor written in eq.
(\ref{sseisextra}). It is not difficult, however, to find the relation between
these two notations. If 
$\epsilon_1$ and $\epsilon_2$ are the two components of the real spinor written
in eq. (\ref{sseisextra}), the complex spinor is simply:
\beq
\epsilon\,=\,\epsilon_1\,+\,i\epsilon_2\,\,.
\label{apados}
\eeq
From eqs. (\ref{sseisextra}) and (\ref{apados}) it is straightforward to find
the following rules to pass from one notation to the other:
\beq
\epsilon^*\,\leftrightarrow\,\sigma_3\,\epsilon\,\,,
\,\,\,\,\,\,\,\,\,\,\,\,\,\,\,\,\,\,\,
i\epsilon^*\,\leftrightarrow\,\sigma_1\,\epsilon\,\,,
\,\,\,\,\,\,\,\,\,\,\,\,\,\,\,\,\,\,\,
i\epsilon\,\leftrightarrow\,-i\sigma_2\,\epsilon\,\,.
\label{apatres}
\eeq
We will use the correspondence written in eq. (\ref{apatres}) to express the
result of our supergravity analysis in the notation (\ref{sseisextra}).

With our notations for the gauge forms, 
the supersymmetry transformations of the dilatino $\lambda$ and gravitino 
$\psi$ in type IIB supergravity are \cite{SUSYIIB}:
\bear
\delta\lambda&=&i\Gamma^{M}\epsilon^{*}\,P_{M}\,-\,{1\over 24}\,
\Gamma^{M_1M_2M_3}\,\epsilon\,F_{M_1M_2M_3}\,\,,\rc\rc
\delta\psi_{M}&=&D_{M}\epsilon\,-\,{i\over 1920}\,\Gamma^{M_1\cdots M_5}
\Gamma_M\,\epsilon\,F^{(5)}_{M_1\cdots M_5}\,+\rc\rc
&&+{1\over 96}\,\Big(\,\Gamma_{M}^{\,\,\,M_1M_2M_3}\,-\,9\,
\delta_M^{M_1}\,\Gamma^{M_2M_3}\,\Big)\,\epsilon^*\,F_{M_1M_2M_3}\,\,.
\label{apacuatro}
\eear
In eq. (\ref{apacuatro}) the $\Gamma^{M}$'s are  ten-dimensional Dirac
matrices with curved indices, $F^{(5)}$ is the RR five-form and 
$P_{M}$ and $F_{M_1M_2M_3}$ are given by:
\bear
P_M&=&{1\over 2}\,\big[\,\partial_M\phi_D\,+\,ie^{\phi_D}\,
\partial_M\chi\,\big]\,\,,\rc\rc
F_{M_1M_2M_3}&=&e^{-{\phi_D\over 2}}\,H_{M_1M_2M_3}\,+\,
ie^{{\phi_D\over 2}}\,F^{(3)}_{M_1M_2M_3}\,\,,
\label{apaccinco}
\eear
where $\chi$ is the RR scalar and $H$ and $F^{(3)}$ are, respectively, the NSNS
and RR three-form field strengths. 

The solutions of the supergravity equations we are dealing with are purely
bosonic and, thus, they are only invariant under those supersymmetry
transformations which do not change the fermionic fields $\lambda$ and $\psi$.
Let us consider first the variation of the dilatino $\lambda$ for the (D1,D3)
background. From eqs. (\ref{uno}), (\ref{cuatro}) and (\ref{trece}) it follows
that the non-vanishing components of the complex three-form $F$ are:
\bear
F_{01r}&=&i\sin\varphi \,\,h_3^{{1\over 4}}\,\partial_r\,f_3^{-1}\,\,,\rc\rc
F_{23r}&=&\sin\varphi \cos\varphi\,\,h_3^{{7\over 4}}\,\partial_r\,f_3^{-1}\,\,.
\label{apaseis}
\eear
By using eq. (\ref{apaseis}), and by computing $P_M$ from eq. (\ref{uno}), one
easily finds from the first equation in (\ref{apacuatro}) that the
supersymmetry variation of $\lambda$ is:
\beq
\delta\lambda\,=\,{\sin\varphi\over 4}\,f_3^{{1\over 2}}\,h_3^{{5\over 8}}
\partial_r\,f_3^{-1}\,\Gamma_{\underline {r}}\,\Big[\,
-i\sin\varphi\,h_3^{{1\over 2}}\,f_3^{-{1\over 2}}\epsilon^*\,+\,
\Gamma_{\underline {x^0x^1}}\epsilon\,+\,i\cos\varphi\,h_3^{{1\over 2}}\,
\Gamma_{\underline {x^2x^3}}\epsilon\,\Big]\,\,.
\label{apasiete}
\eeq
Clearly $\delta\lambda\,=\,0$ if and only if the term inside the brackets on the
right-hand side of eq. (\ref{apasiete}) vanishes. It is an easy exercise to
verify, by using the correspondence (\ref{apatres}),  that the condition so
obtained coincides with the $\kappa$-symmetry condition (\ref{stdos}). Thus, we
can represent the spinor $\epsilon$ as in eq. (\ref{stcuatro}) with $\alpha$
given in eq. (\ref{sttres}) and $\tilde \epsilon$ satisfying (\ref{stcinco}).

More information about the spinors which leave invariant the (D1,D3) solution
can be gathered by looking at the gravitino transformation rule
(\ref{apacuatro}). Let us consider first the components of $\psi$ along the
directions $x^{\mu}$ ($\mu=0,1,2,3$) parallel to the D3-brane. Due to the
presence of a covariant derivative on the expression of $\delta\psi_M$, in order
to obtain the variation of the gravitino, we  need to know the value of the spin
connection. It is straightforward to check that the only non-vanishing
components of the latter of the type 
$\omega_{x^{\mu}}^{\underline{MN}}$ are:
\bear
\omega_{x^0}^{\underline{x^0 r}}&=&\omega_{x^1}^{\underline{x^1 r}}\,=\,
{1\over 4}\,f_3^{{1\over 2}}\partial_r
f_3^{-1}\,+\,{1\over 8}\,\sin^2\varphi\,h_3\,f_3^{-{1\over 2}}\partial_r
f_3^{-1}\,\,,\rc\rc
\omega_{x^2}^{\underline{x^2 r}}&=&\omega_{x^3}^{\underline{x^3 r}}\,=\,
{1\over 4}\,f_3^{{1\over 2}}\,h_3^{{1\over 2}}\partial_r
f_3^{-1}\,-\,{3\over 8}\,\sin^2\varphi\,h_3^{{3\over 2}}
\,f_3^{-{1\over 2}}\partial_r
f_3^{-1}\,\,.
\label{apaocho}
\eear
By using  eq. (\ref{apaocho}), the values of the forms given in eqs. 
(\ref{trece}) and (\ref{apaseis}) and the constraint imposed on $\epsilon$ by
the invariance of the dilatino (eq. (\ref{stdos})), one concludes after some
calculation that the gravitino components   $\psi_{x^{\mu}}$ ($\mu=0,1,2,3$)
are invariant under supersymmetry if and only if the Killing spinor is
independent of the $x^{\mu}$'s, namely:
\beq
\partial_{x^{\mu}}\,\epsilon\,=\,0\,\,,
\,\,\,\,\,\,\,\,\,\,\,\,\,
(\mu=0,1,2,3)\,\,.
\label{apanueve}
\eeq

The condition $\delta\psi_{\rho}\,=\,0$ determines the dependence of $\epsilon$
on $\rho$. The relevant component of the spin connection needed in the
calculation of $\delta\psi_{\rho}$ is:
\beq
\omega_{\rho}^{\underline{\rho r}}\,=\,
{1\over \sqrt{1\,-\,\rho^2}}\,\,\Bigg[\,
1\,-\,{1\over 4}\,r\,f_3\,\partial_r\,f_3^{-1}\,+\,
{1\over 8}\,r\,\sin^2\varphi\,h_3\,\partial_r\,f_3^{-1}\,\Bigg]\,\,.
\label{apadiez}
\eeq
(Notice that when $r\rightarrow\infty$ we recover eq. (\ref{stsiete})). Using
again the condition imposed by the dilatino invariance, one can easily prove
that the dependence on $\rho$  of $\epsilon$ is the same as in eq. 
(\ref{stocho}). Thus, we have verified the representation of $\epsilon$ in
terms of the spinor $\hat\epsilon$ written in eq. (\ref{stcuatro}) and 
(\ref{stocho}). An explicit representation of $\hat\epsilon$ can be obtained by
looking at the transformation of the other components of the gravitino. For
instance, one can consider the equation for the radial component of $\psi$.
Using the fact that $\omega_{r}^{\underline{MN}}\,=\,0$, one gets that the
dependence of $\hat\epsilon$ on $r$ is determined by the  equation:
\beq
\partial_r\hat\epsilon\,=\,-{1\over 8}\,\partial_r\,
\Big[\,\ln\big(f_3h_3^{{1\over 2}}\big)\,\Big]\,\,\,\hat\epsilon\,\,,
\label{apaonce}
\eeq
which can be immediately integrated. One can proceed similarly with the
remaining components of $\psi$. The final result of this analysis is the
complete determination of the form of $\hat\epsilon$ and, therefore, of the
complete Killing spinor $\epsilon$.  
One gets:
\bear
\epsilon\,&=&\,\Big[\,f_3\,h_3^{1/2}\,\Big]^{-{1\over 8}}\,\,
e^{{1\over 2}\,\alpha\,\Gamma_{\underline {x^2x^3}}\,\sigma_3}\,\,
e^{-{1\over 2}\,\beta\,\Gamma_{\underline {\rho r}}}\,\,\,
e^{-{1\over 2}\,\phi\,\Gamma_{\underline {\phi r}}}\,\,\,\times\rc\rc
&&\times\,e^{-{1\over 2}\,\theta_1\,\Gamma_{\underline {\theta_{1}\rho}}}\,\,
e^{-{1\over 2}\,\theta_2\,\Gamma_{\underline {\theta_{2}\theta_{1}}}}\,\,
e^{-{1\over 2}\,\theta_3\,\Gamma_{\underline {\theta_{3}\theta_{2}}}}
\,\,\epsilon_0\,\,,
\label{apadoce}
\eear
where $\epsilon_0$ is a constant spinor satisfying the condition
\beq
(i\sigma_2)\Gamma_{\underline{x^0\cdots x^3}}\,\epsilon_0\,=\,
\epsilon_0\,\,.
\label{apatrece}
\eeq
It follows from eqs. (\ref{apadoce}) and (\ref{apatrece}) that the 
(D1,D3) background is ${1\over 2}$ supersymmetric.

\end{document}